# A Real Time Lighting Technique for Procedurally Generated 2D Isometric Game Terrains

Érick O. Rodrigues and Esteban Clua

Department of Computer Science, Universidade Federal Fluminense,
Rua Passo da Pátria 156, Niterói - RJ, Brazil
erickr@id.uff.br, esteban@ic.uff.br

**Abstract.** This work proposes an automatic real time lighting technique for procedurally generated isometric maps. The scenario is generated from a string seed and the proposed lighting system estimates the geometrical shape of the 2D objects as if they were 3D for further light interaction, therefore producing a 2.5D effect. We employ opacity maps to overcome an issue generated by the geometrical shape estimation. The solution is a coupled approach between the CPU and GPU. The produced visuals, gameplay and performance were evaluated by gamers, programmers and designers. Furthermore, the performance, in terms of frames per second, was evaluated over distinct graphics cards and processors and was satisfactory.

## 1 Introduction

As the processing power of computers improves over time, so does the opportunity for more complex game architectures and mechanics. Procedural Content Generation (PCG) techniques for games enable the construction of more immersive, defying, lasting and realistic games. The game Spore, for instance, employs several concepts of evolution, where creatures, textures and game spaces are procedurally generated to simulate natural selection, evolution and exploration of an infinite universe. In addition, PCG has the feasibility to reduce manual design efforts of nearly every part of the game and the time required for game development [1] while having the potential of contributing greatly to the reduction of the game data.

Shadows and lights in games are frequently computed on the basis of 3D geometries of scene objects. Although isometric terrains produce a visually pleasing result, resemble a 3D world, and conform to the effort reduction in game production due to disregarding 3D models and working with 2 dimensions instead, such as in [2] and as shown in Figure 1, their illumination is not trivial due to the lack of geometrical data. A possible solution for this is the usage of normal maps or even a simplistic volume information associated to the 2D scene that is employed as a guide for light and shadow interaction [3,4]. However, these kind of solutions require dedicated design of the scene elements, which in many cases may be impracticable.

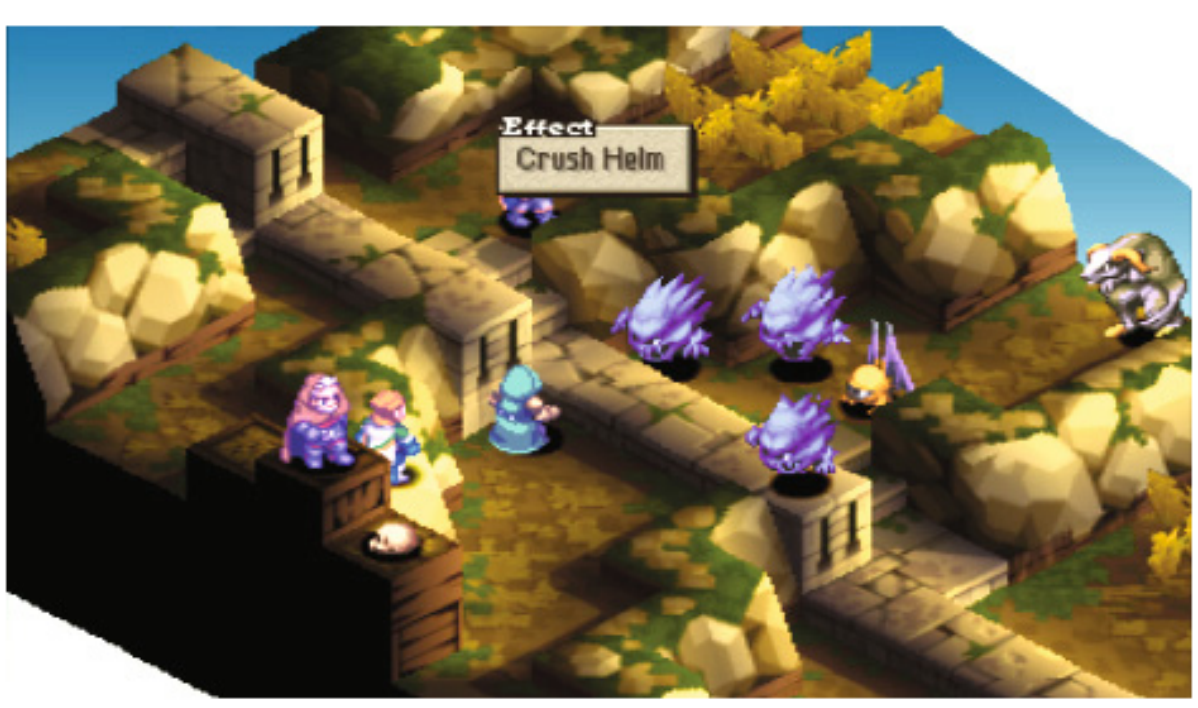

**Fig. 1.** Final fantasy tactics, one of the most famous games that adopts the isometric perspective.

This work proposes (1) a novel technique for estimating the geometry of 2D sprites in the isometric environment and (2) an approach for tracing lights and shadows of the scene based on the CPU and GPU. Both of these steps are automatically generated. In other words, given a certain scene having several distinct 2D sprites and no access to any kind of geometrical information nor normal maps, we propose a lighting system that interacts with the 2D environment and produces 3D-like illumination.

## 2 Literature Review

Currently, the literature on PCG is scattered across numerous fields. However, it is primarily related to Computer Science in areas such as Computer Graphics, Pattern Recognition, Games, Artificial Intelligence and Multimedia Computing. The earliest games that employed PCG were produced around 1980. The exploration game Elite is one of the earliest and employed a Pseudo-Random Number Generation (PRNG) to produce a very large universe [5]. When applying PRNG, the data is generated using a seeded algorithm, which allows the process to be deterministic. This is an important feature since the automatically generated data can be consistently reproduced and therefore tested [6]. Perlin Noise, for instance, is a PRNG developed to make computer generated images look more realistic [7].

PCG is constantly being applied to a large amount of commercial games. Rogue is a dungeon crawling game of the 1980s. Unlike adventure games of the time, Rogue randomly generated the dungeon layout and the location of items. Furthermore, PCG was applied to a couple of games at the time such as Telengard, Nethack and Elite but further nearly vanished from the mainstream of commercial games. PCG was easy to implement in games of this period, which run over DOS. As soon as the frameworks became more complex, the appliance of PCG became even more complex and was consequently rarely regarded.

In 1996, Diablo resurrected PCG and roguelike games to the mainstream of commercial games. Diablo procedurally generates its levels and loot. Its sequel, Diablo II, maintained the PCG characteristic of the former. Every time a player starts the game, the maps and levels are assembled distinctly. Both Diablo and Diablo II are 2D games in the isometric perspective. The first Diablo game featured no lighting effects apart from what can be done with sprites. Diablo II, on the other hand, used a simplistic pre-built geometry of the scene to apply an incipient 2D illumination.

Randomness is one of the basis of PCG. In general, what mainly changes amongst several approaches is the chosen threshold of randomness and the way it is applied to the problem. Evolutionary algorithms, which involve random mutations, are often applied to the procedural generation of characters [8], terrains [9], tracks for racing games [10] and others. However, as previously addressed, deterministic algorithms can also be classified as procedural. Although the result of the computation is always the same for a combination of input parameters, the set of parameters itself (seed) may be randomly generated. In this work, the generated lights and shadows vary related to the position of the light and to the environment in a deterministic fashion. Therefore, the maps, as well as the positions, are the seed of the algorithm.

Ebert et al. [11] introduced several methods for procedurally generating game contents. Among these methods, algorithms for generating solids, gases, water, fire, noise, cloud, earth textures and materials, for instance, were regarded. Kelly et al. [12] surveyed techniques for procedurally generating cities, focusing on individual buildings, road networks and cityscapes. Smelik et al. [13] surveyed procedural generation of terrain and urban environments. At last, Hendrikx et al. [5] surveyed several aspects of PCG for games and categorized the existing algorithms in 6 main distinct classes: Game Bits, Game Space, Game Systems, Game Scenario, Game Design and Derived Content. The class (1) Game Bits comprises the generation of texture, sound, vegetation, buildings, behavior, fire, water, stone and clouds; (2) Game Spaces, on the other hand, aggregates indoor maps (e.g., rooms in general), outdoor maps (e.g., terrains) and bodies of water (e.g., river, lakes and seas). Furthermore, ecosystems, road networks, urban environments and entity behavior are comprised by the (3) Game Systems class. Notwithstanding, puzzles, storyboards, story and levels are fit in the (4) Game Scenario class. If the algorithm refers to the creation of mathematical patterns underlying the game and game rules then it was categorized as (5) Game Design. Finally, news and leaderboards are fit in the (6) Derived Content class.

The approach proposed in this work does not fit properly in any of the classes defined by Hendrikx et al. [5]. That is so because we are the first to introduce a methodology that procedurally estimates the geometry of 2D sprites for further light interaction. In our methodology, we draw the light rays to a texture and further pass it to the fragment shader. Thus, it could be partially comprised by the Game Bits category. However, since the entire methodology does not just generate a texture, it appears to be misclassified. Perhaps a new class called Game Effects could be added to the definition of Hendrikx et al. for a proper categorization.

## 3 The Approach

The map architecture followed the Entity Component System (ECS) definition. That is, every graphical component of the game is an extension of the Entity class that has a $(x, y)$ position property. In this work we propose the usage of auxiliary layers that overlap each other. Thus, each tile or block of the map, as shown in Figure 2, is composed of the block sprite plus at most 3 layers of overlapping sprites such as shown in the third image of Figure 2.

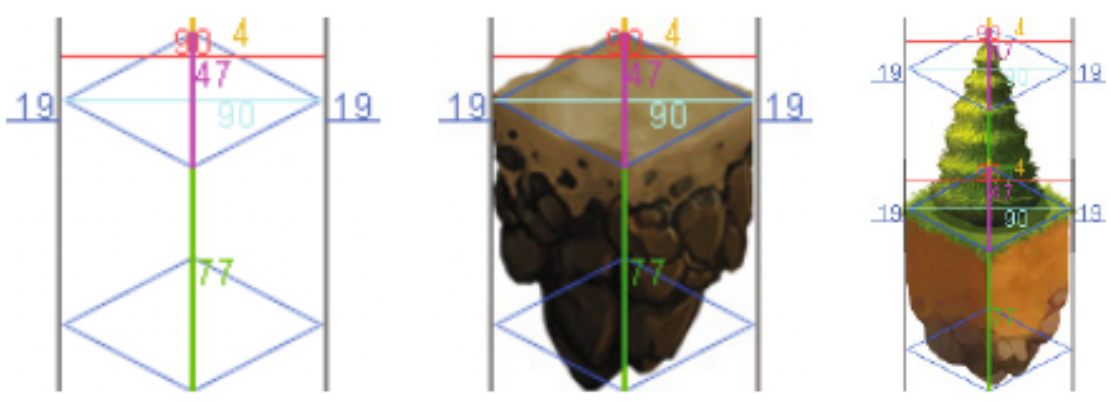

**Fig. 2.** Tiles or blocks of the map.

The blocks of the map are then assembled together. The assemblage is straightforward: the $n^{th}$ block is placed at the right position of the $(n-1)^{th}$ block, as long as there is no line break. If a line break is present, assuming that the first block of the previous line is called $p$, then the $n$ block is placed at the bottom of $p$ in the $y$ direction, and at the center of $p$ and $p+1$ in the $x$ direction. A randomly generated map is shown in Figure 3.

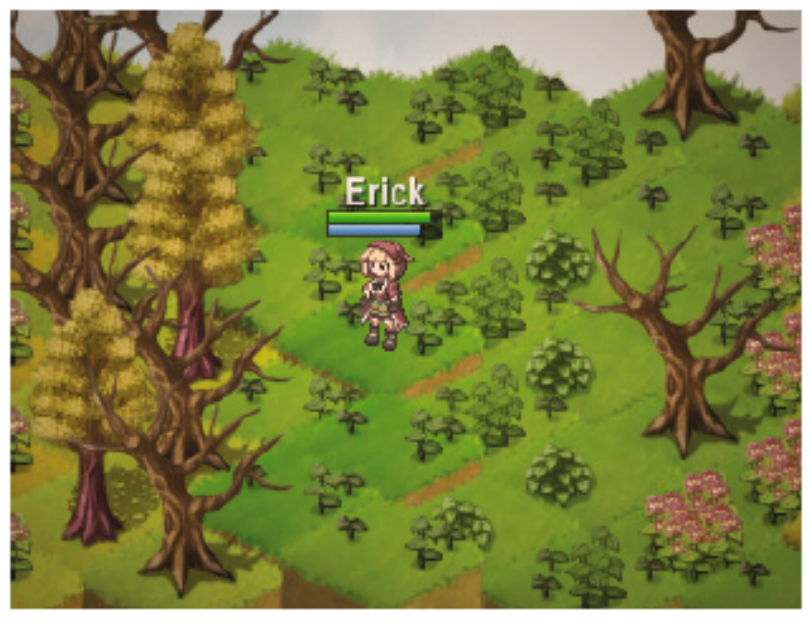

**Fig. 3.** Randomly generated map.

The approach consists of, at first, creating an obstacle map every time a new map is loaded and storing it in memory. At every iteration of the render method, the light sources that appear within the scene are processed and drawn to a texture of slightly greater size than the screen. The light rays are traced using the Bresenham's Line Algorithm [14]. The produced texture is then bound to the fragment shader and the fragment shader alters the colors of each fragment according to the light texture. The entire process is illustrated in Figure 4.

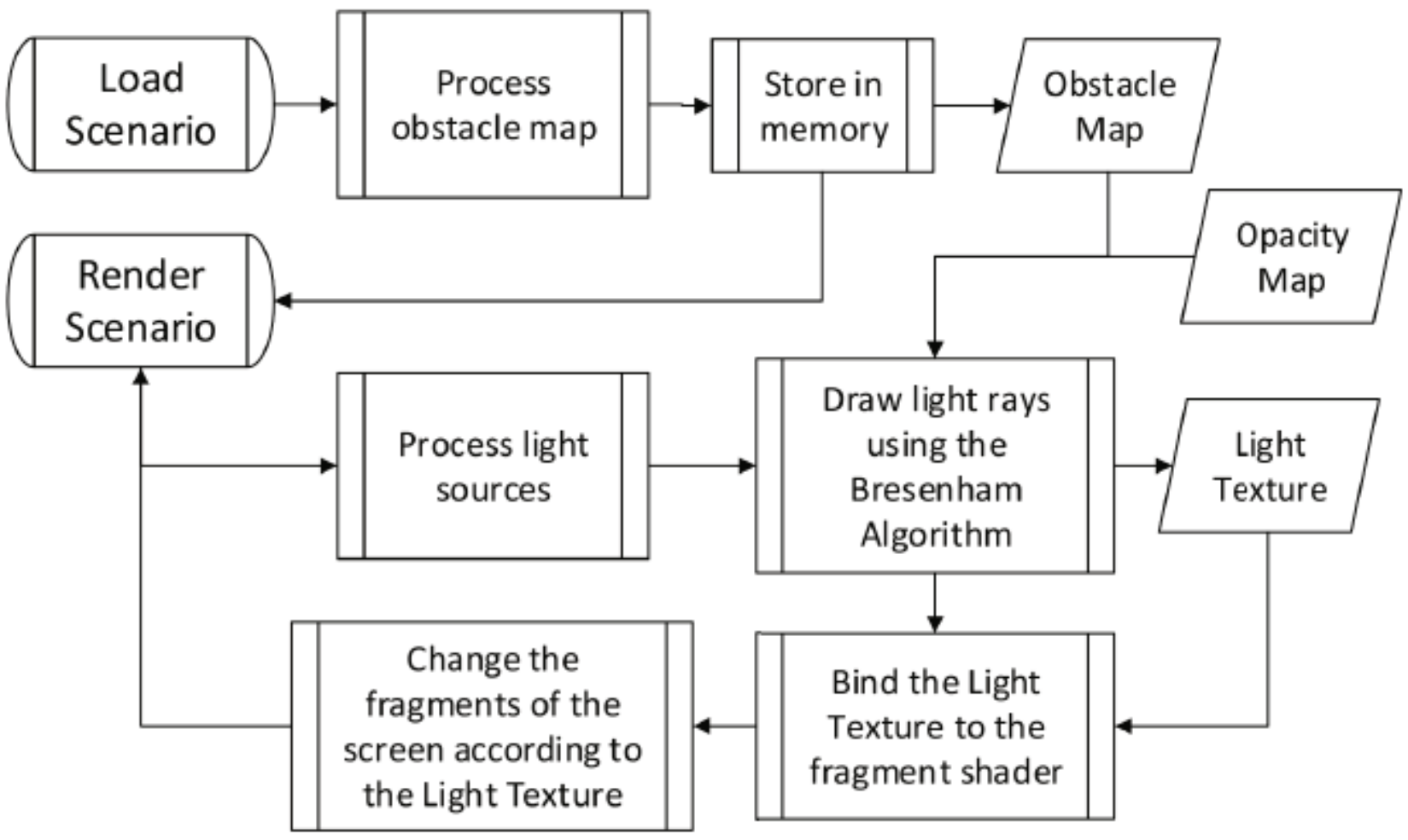

**Fig. 4.** Overall steps of the proposed approach.

### 3.1 Obstacle Map

As previously addressed, when a new map is loaded, a 2D boolean obstacle map is created to store information that is used to halt the tracing of light rays if they hit an opaque object. A heuristic was used to compute the obstacle map and is shown in Algorithm 1. Figure 5 illustrates the $\Delta y$ of an arbitrary block. The height of $\Delta y$ was empirically chosen to be 25 pixels in our case, where each sprite of the block is 128x128 pixels wide.

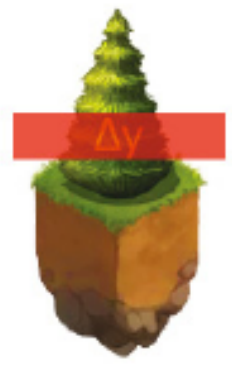

**Fig. 5.** The $\Delta y$ in an arbitrary block.

**while** *not every block b has been iterated* **do**
  1. read $\Delta y$ lines of $b$ and store the largest sequence of non-transparent pixels for each line;
  2. take the sequence of occurrences $s$ that has the least deviation to the mean;
  3. set the pixels of $s$ as true in the obstacle map;
  4. take the central pixel of $s$ and set their upper and lower $s.length/4$ pixels of the obstacle map as true;
**end**

**Algorithm 1.** Constructing the obstacle map.

Thereafter, a box-shaped noise reduction was applied to erase any tiny obstacle from the obstacle map. That is, we consider a $n \times n$ window and displace this window over the obstacle map verifying if there is less than $n$ pixels set as obstacle within the window. If so, the top-left pixel of the window is unset as obstacle.

### 3.2 Light Rendering

Once at each $l$ frames, the light rays are traced, interacting with the obstacle map. The Bresenham Algorithm [14] was used to trace these lines and was properly adapted to the problem as shown in Algorithm 2. At every light source within the screen a surrounding rectangular area of size $a \times b$ is regarded for tracing the light rays. A total of $4ab$ light rays are traced towards every pixel of the border of this area from the central pixel of each light source.

The Algorithm 2 receives the coordinates of the light source's origin and the coordinates of every pixel of the rectangular border. The pixels corresponding to the traced rays are drawn in a particular order: from the center of the light source to the pixels of the rectangular border. During this process, if any of these pixels corresponds to an obstacle pixel in the obstacle map, then the tracing halts at this specific pixel. If no obstacle is found, the algorithm draws the intensity and transparency of the light texture's pixel based on the function $i(\Delta j, \Delta i) = 255/(1 + max(\Delta j, \Delta i))$, where $\Delta j$ and $\Delta i$ represent the distance of the iterated pixel with regard to the central pixel of the light source and assuming that the texture is 8-bits depth (max value: 255).

It is interesting to note that not necessarily a light ray must be traced for every pixel of the rectangular border. Some of these pixels can be skipped to improve the overall performance and the eventual produced gaps can be further corrected in the shader at the GPU with some kind of blurring algorithm [15]. Furthermore, we previously defined that the light rays are traced at every $l$ frames. Thus, while the light texture is not updated among these $l$ frames, the texture should be displaced (inverse translation) according to the main central moving character, to avoid incorrect placements through the screen of a light that was generated at a certain position at a certain frame. Finally, if the generated light texture is directly drawn to the screen, then the result would look like

```
method traceLightRay(int orgX, int orgY, int dstX, int dstY);
begin
    int w = (dstX - orgX), h = (dstY - orgY);
    short dx1 = 0, dy1 = 0, dx2 = 0, dy2 = 0;
    if (w < 0) dx1 = -1; dx2 = -1; else if (w > 0) dx1 = 1; dx2 = 1;
    if (h < 0) dy1 = -1; else if (h > 0) dy1 = 1;
    short longest = absolute(w); short shortest = absolute(h);
    if !(longest > shortest) then
        longest = absolute(h); shortest = absolute(w); dx2 = 0;
        if (h < 0) dy2 = -1; else if (h > 0) dy2 = 1;
    end
    int numerator = longest >> 1; boolean finished = false;
    for int i=0; i ≤ longest and !finished; i++ do
        if pixel(orgX, orgY) is not blocked then
            enlighten pixel(orgX, orgY);
        else
            finished = true;
        end
        numerator += shortest;
        if !(numerator < longest) then
            numerator -= longest;
            orgX += dx1; orgY += dy1;
        else
            orgX += dx2; orgY += dy2;
        end
    end
end
```

**Algorithm 2.** Adapted Bresenham Algorithm.

Figure 6. The white pixels on top of the trees and bushes represent the obstacle map.

The arrows in Figure 6 indicate issues generated by the current approach. That is, the bushes that are at a lower position in the vertical direction than the light's center should be completely dark, not half enlightened as it currently is. Furthermore, the part of the tree pointed by the red arrow should have been enlightened, but the current approach projects a shadow in this area.

To overcome this issue we introduce the concept of opacity maps. Each sprite of the scene (excluding the floor blocks) has its own opacity map. An opacity map is a boolean matrix that indicates if a certain pixel of the sprite is transparent or not, given a certain threshold. Thus, the opacity maps of the sprites are subtracted or added to the light texture. If the iterated sprite is at a higher position than the center of the light in the vertical direction, the opacity map of the same is added to the light texture and their illumination coefficient are properly computed. Otherwise, if the sprite is at a lower position, then its opacity map is subtracted from the light texture. The result of this correction is shown in Figure 7.

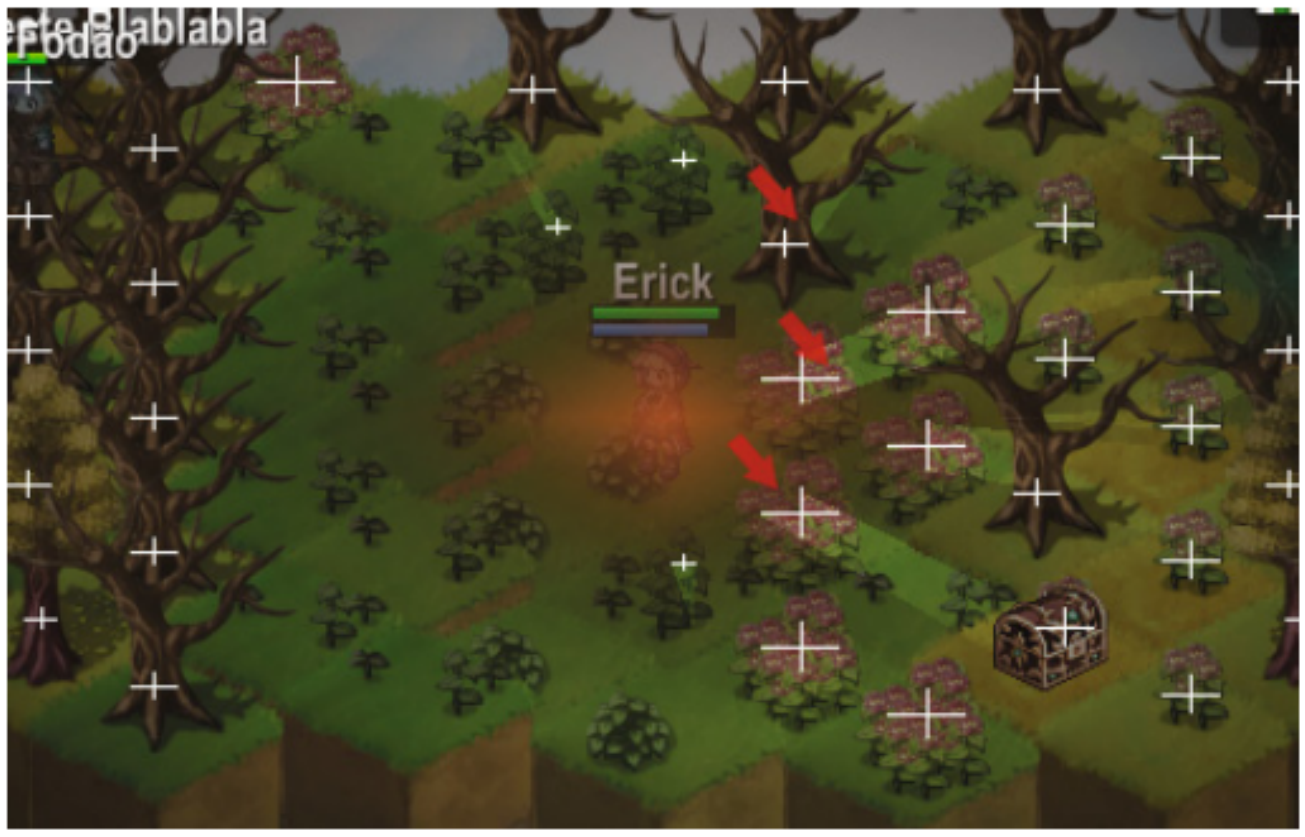

**Fig. 6.** Light texture being directly rendered on top of the sprites.

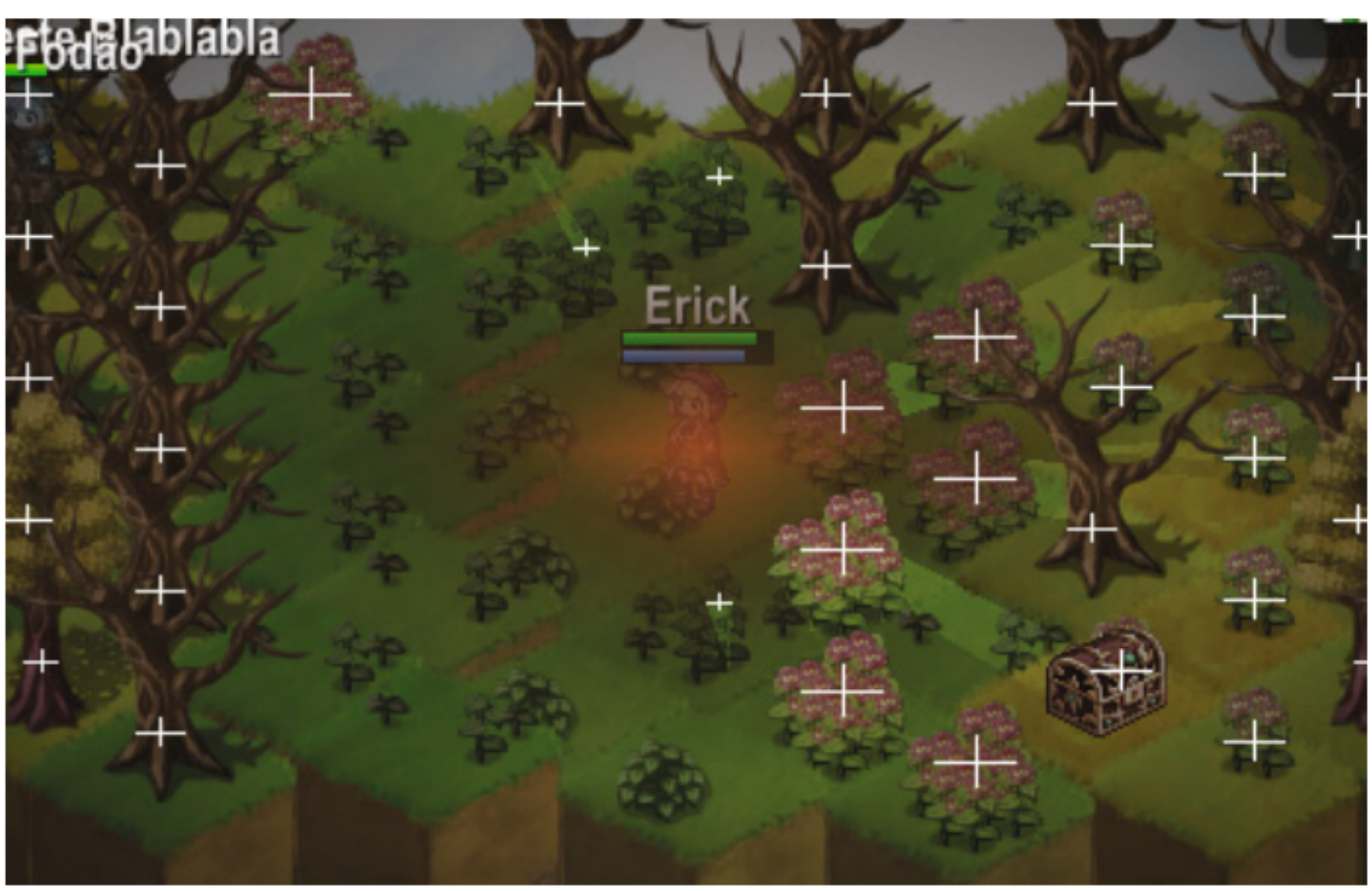

**Fig. 7.** Opacity maps added to the light texture.

After modifying the light texture according to the opacity maps of the surrounding sprites, the texture is bound to the shader and the color of the fragments are computed following the Algorithm 3. Essentially, the GPU will just receive the light texture and change the colors of the fragments according to the previously bound texture. Moreover, in our implementation, the framebuffer was also used for drawing all the scene sprites priorly to applying the lighting effects. Once the light texture is in the GPU, a simple box-blur algorithm is applied to the same. Although the computations in the GPU are an essential part of the process, the method is much more associated to the CPU. Therefore, processors with shared memory between the GPU and CPU will benefit from the approach.

**Data**: texColor being the color of the fragment, ambientClarity being a float that stores the ambient clarity and lightTexture being the texture binded to the shader

```
begin
    (...);
    float luminance = texColor.r*0.349 + texColor.g*0.114 + texColor.b*0.537;
    vec4 lightFactor =
    luminance*lightTexture/(ambientClarity*ambientClarity+0.1);
    vec4 texFactor = texColor*sqrt(ambientClarity);
    texColor = texFactor + lightFactor;
end
```

**Algorithm 3.** Fragment shader algorithm.

## 4 Results

Figure 8 shows the final result with two light sources on both characters [16]. The ambient clarity in this occasion was set to 0.4. The blurring gives a final nice aesthetics and allows the possibility of skipping some light rays during the tracing to speed up the overall performance. If the texFactor variable in Algorithm 3 is multiplied by lightTexture, then what is not hit by the light texture would appear completely dark.

Furthermore, we do not specifically generate shadows. That is, we assume that every pixel that does not have an associated light ray is a shadow. However, if one desires to cast shadows instead of light rays, then the light ray tracing just need to be inverted. In other words, at its current state, the light rays are traced until they reach an obstacle. To cast shadows, the shadow rays should be painted after the light rays reach the same obstacle.

The achieved Frame Rate Per Second (FPS) was satisfactory for a procedurally generated approach. In some computers the drop in FPS was not noticeable. Figure 9 illustrates some of the obtained FPS among distinct processors and graphics cards. It is interesting to notice that, in fact, the bottleneck of the approach is at the CPU and at the communication between the CPU and GPU. The processors that have an integrated GPU obtained better results due to memory bandwidth issues. The overall power of the GPU alone did not influence positively on the obtained FPS.

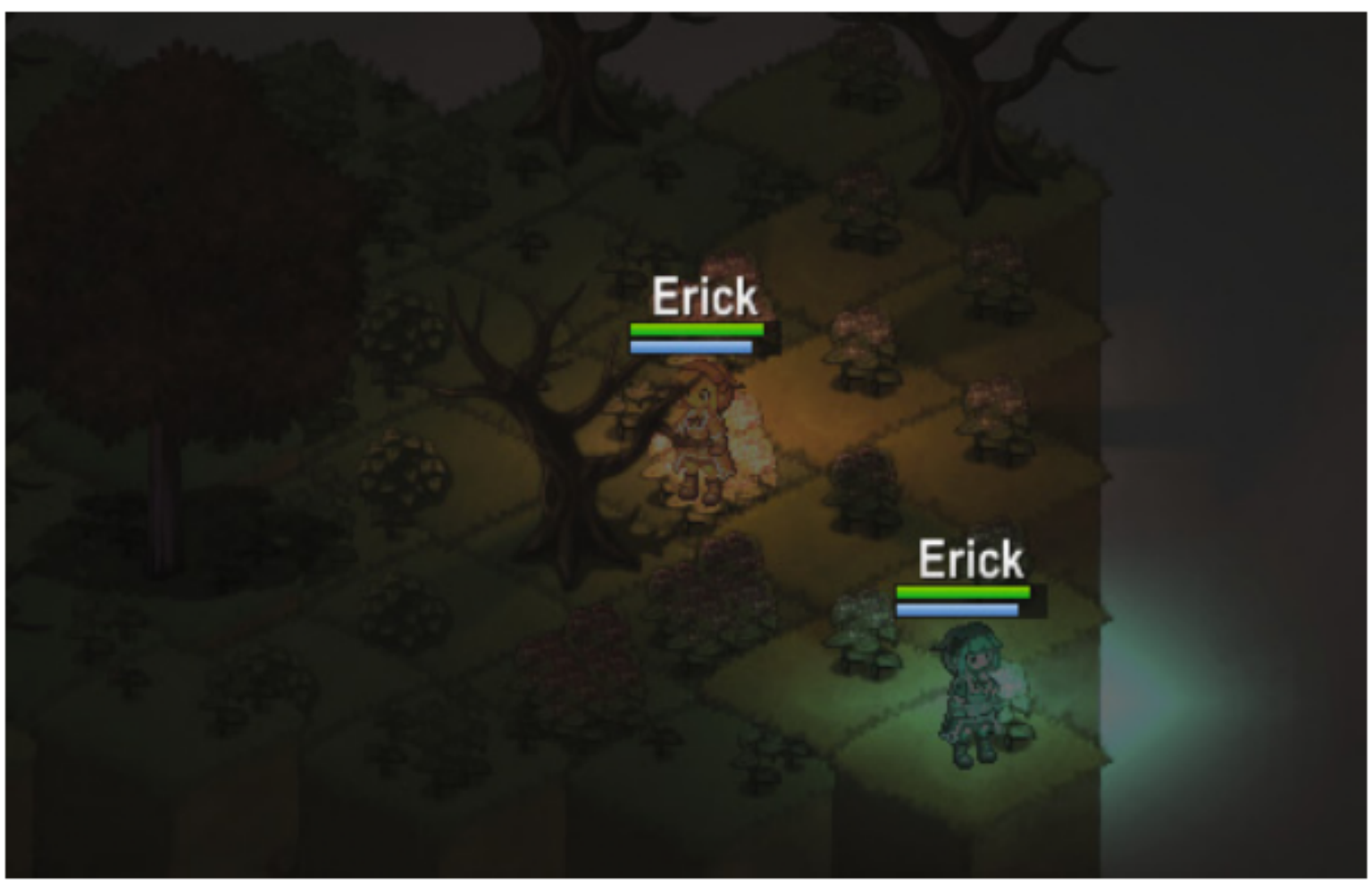

**Fig. 8.** Visual result.

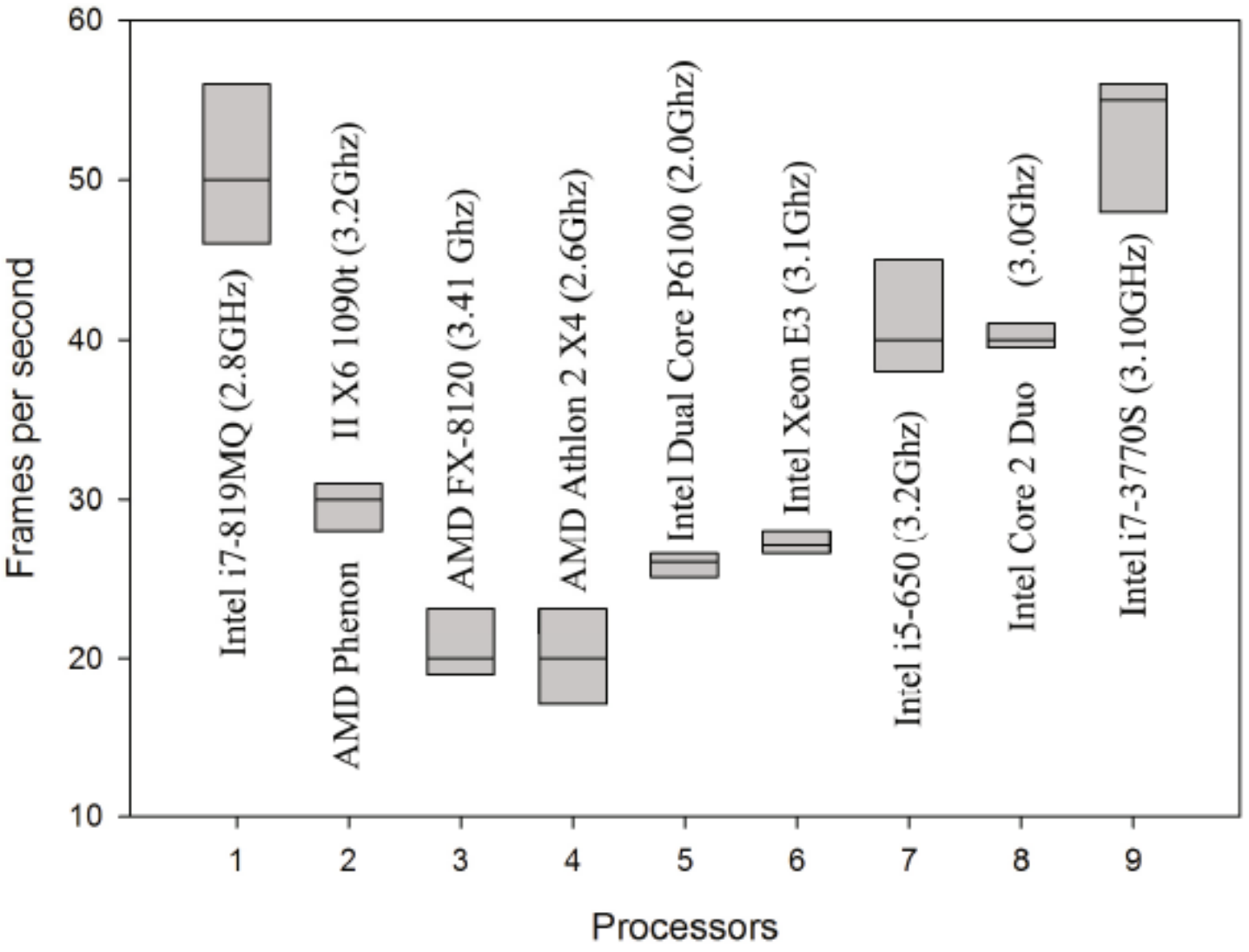

**Fig. 9.** A benchmark of the approach.

The graphics cards used in each case from 1 to 9 were, respectively: (1) Nvidia GeForce GTX 850M,(2) Nvidia GeForce GTX 550ti, (3) Nvidia GeForce 630, (4) ATI Radeon HD 5550, (5) Intel HD Graphics, (6) ATI Radeon 280X Windforce, (7) Intel HD Graphics, (8) Nvidia GeForce GT 440 and (9) Intel HD Graphics. It is clear that the Intel processors performed in a more efficient fashion than AMD, specially in the chips that have an integrated graphics card. In addition, it appears that when processors and graphics cards are of the same manufacturer, the FPS is slightly benefited such as on case 4 if compared to 3. However, we cannot assure that correlation.

Beyond that analysis, we have also collected feedbacks from 61 evaluators including gamers, designers and programmers regarding the usability, benefits and visuals produced by the approach. Among them 40 were designers, 55 gamers and 42 programmers. Their analysis are summarized in Table 1. The last concept of the table is related to a comparison between the proposed approach and a non-interactive approach, such as the one in image (b) in Figure 10 (including movement, not only static). It is important to highlight that the resultant colors between (a) and (b) are a little different, though we have tried to minimize that difference as much as possible.

**Table 1.** General concepts analysis.

| Concept | Designers | Gamers | Programmers | Mean |
|---|---|---|---|---|
| Overall rating (0 to 10) | 8.72 | 8.74 | 8.85 | 8.65 |
| Possible game production speed up rating (0 to 10) | 9.50 | 9.27 | 9.28 | 9.18 |
| Would active the lighting with no noticeable FPS drop (%) | 97.5 | 85.4 | 90.4 | 85.2 |
| Would active the lighting even with noticeable FPS drop (%) | 67.5 | 58.1 | 66.6 | 60.6 |
| Would reduce resolution to activate the lighting (%) | 47.5 | 38.2 | 40.4 | 39.3 |
| Find it better than no interaction, Figure 10-(b) (%) | 86.5 | 80.3 | 80.1 | 79.4 |

## 5 Conclusion

The obtained results with our proposed approach are satisfactory in the visual and performance aspects. We have shown that the inclusion of automatic illumination in procedurally generated isometric scenes can be achieved without the need of proper designing or modelling. Furthermore, we have also shown that our method is independent of the sprites topology shapes.

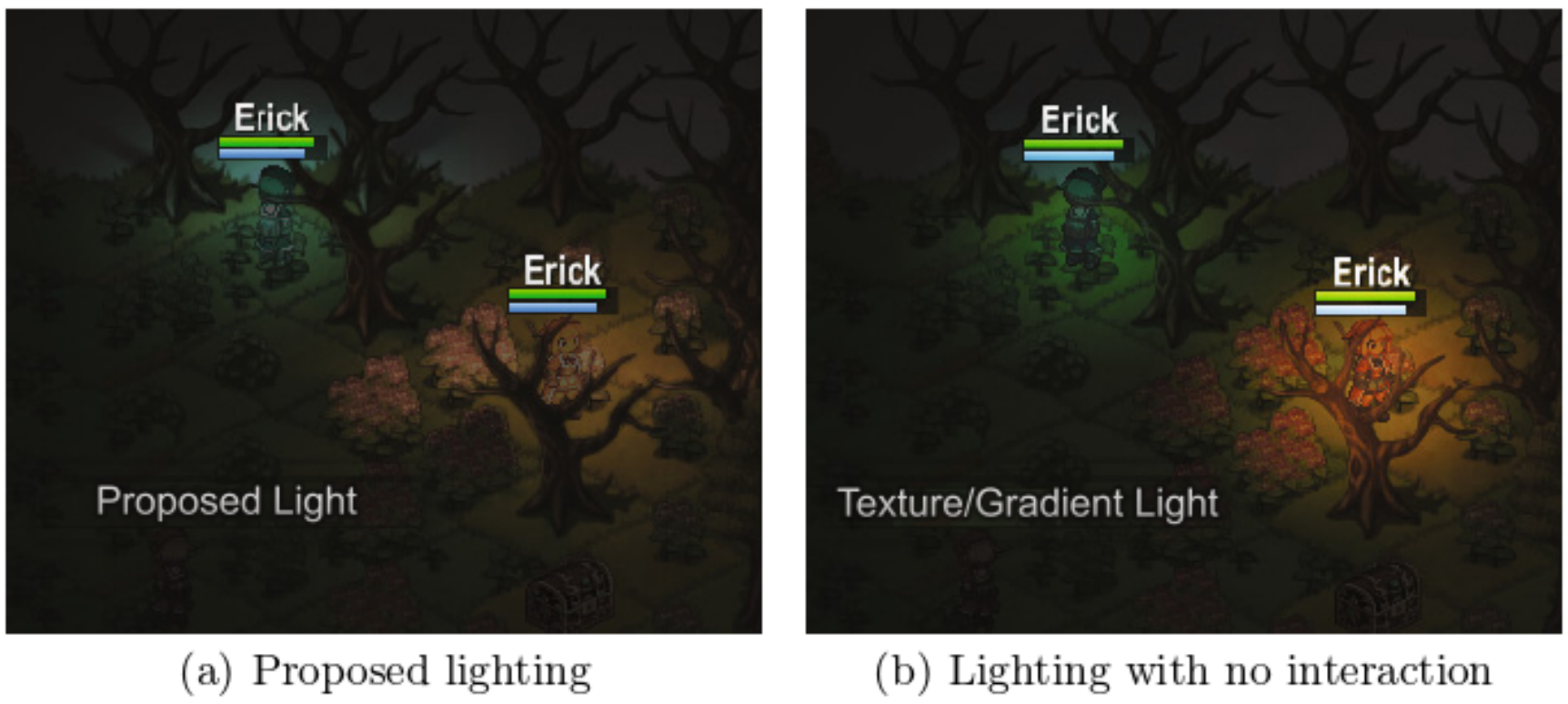

(a) Proposed lighting (b) Lighting with no interaction

**Fig. 10.** Comparison of the light interaction.

Moreover, the approach may contribute as an architectural basis for a framework that can procedurally generate isometric maps of the addressed configuration. Due to the fact that the geometrical shapes of the objects on the scene and the lighting interaction are procedurally generated, there is no need to be concerned with anything other than the seed generation, which involves no 3D modelling and, therefore, greatly diminishes the workload. As a future work, we intend to develop a sandbox framework, making it possible to include in different tools.

We can conclude from the collected feedback that the approach is very interesting from the perspective of the user, especially on the PCG-related aspects. The greatest ratings were on the possibility of speeding up the production of games if the proposed lighting is regarded. However, we can also conclude that the majority would not use or activate the proposed lighting if the drop in FPS is such that it affects the gameplay experience. This argument came in a more accentuated fashion from gamers than from developers (designers and programmers). Fortunately, with regard to some computers specifications, the proposed lighting produces no apparent FPS drop. As future work we would also like to improve the CPU bottleneck.

## References

1. Lee, Y.-S.: Context-aware petri net for dynamic procedural content generation in role-playing game. Computational Intelligence Magazine 6, 16–25 (2011)
2. Doytsher, Y., Hall, J.K.: Simplified algorithms for isometric and perspective projections with hidden line removal. Computers and Geoscience 27, 77–83 (2001)
3. Kilgard, M.J.: A practical and robust bump-mapping technique for today's gpus. In: Game Developers Conference: Advanced OpenGL (2000)
4. DesLauriers, M.: Normal mapping. `https://github.com/mattdesl/lwjgl-basics/wiki/ShaderLesson6`

5. Hendrikx, M., Meijer, S., Van Der Velden, J., Iosup, A.: Procedural content generation for games: A survey. ACM Transactions on Multimedia Computing, Communications, and Applications, 9 (2013)
6. Smith, A.J., Bryson, J.J.: A logical approach to building dungeons: Answer set programming for hierarchical procedural content generation in roguelike games. Proceedings of the 50th Anniversary Convention of the AISB (2014)
7. Perlin, K.: An image synthesizer. In: Proceedings of the 12th Annual Conference on Computer Graphics and Interactive Techniques, vol. 19, pp. 287–296 (1985)
8. Collazo, M.N., Cotta, C., Fernndez-Leiva, A.J.: Virtual player design using self-learning via competitive coevolutionary algorithms (2014)
9. Raffe, W.L., Zambetta, F., Li, X.: A survey of procedural terrain generation techniques using evolutionary algorithms. Proceedings of Congress of Evolutionary Computation 10, 2090–2097 (2012)
10. Loiacono, D., Cardamone, L., Lanzi, P.L.: Automatic track generation for high-end racing games using evolutionary computation. IEEE Transactions on Computational Intelligence and AI in Games 3, 245–259 (2011)
11. Ebert, D.S., Musgrave, F.K., Peachey, D., Perlin, K., Worley, S.: Texturing and Modeling: A Procedural Approach, 3rd edn. Morgan Kaufmann Publishers (2003)
12. Kelly, G., McCabe, H.: A survey of procedural techniques for city generation. ITB Journal, 14 (2006)
13. Groenewegen, S.A., Smelik, R.M., de Kraker, K.J., Bidarra, R.: Procedural city layout generation based on urban land use models. In: Proceedings of the 30th Annual Conference of the European Association for Computer Graphics, pp. 45–48.
14. Bresenham, J.E.: Algorithm for computer control of a digital plotter. IBM Systems Journal 4, 25–30 (1965)
15. Waltz, F.M., Miller, J.W.V.: Efficient algorithm for gaussian blur using finite-state machines. In: Proceedings of SPIE 3521, Machine Vision Systems for Inspection and Metrology VII (1998)
16. Rodrigues, E.O.: 2d shader light and shadow system, `https://www.youtube.com/watch?v=jpmRXUH2qFU`